\documentclass[fleqn,usenatbib,useAMS]{mnras}

\usepackage{graphicx}	
\usepackage{amsmath}	
\usepackage{amssymb}	
\usepackage{multicol}   
\usepackage{bm}		
\usepackage{pdflscape}	
\usepackage{xspace}

\newcommand{\paperkostasalt}{Karpouzas et al. (2020b, submitted)\xspace}

\newcommand{\MAXI}{\mbox{MAXI J1348--630}\xspace}

\newcommand{\NICER}{\textit{NICER}\xspace}
\newcommand{\XSPEC}{\textsc{xspec}\xspace}
\newcommand{\emcee}{\textsc{emcee}\xspace}
\newcommand{\dual}{\textit{dual}\xspace}

\usepackage[T1]{fontenc}
\usepackage{ae,aecompl}

\usepackage{newtxtext,newtxmath}

\title[A Comptonisation model for the type-B QPO in MAXI J1348--630]{A two-component Comptonisation model for the type-B QPO in MAXI J1348--630}

\author[Federico Garc\'ia et al.]{Federico Garc\'ia$^{1}$%
\thanks{Contact e-mail: \href{mailto:garcia@astro.rug.nl}{garcia@astro.rug.nl}},
Mariano M\'endez$^{1}$, 
Konstantinos Karpouzas$^{1}$, 
Tomaso Belloni$^{2}$, \and
Liang Zhang$^{3}$,
Diego Altamirano$^{3}$
\\
$^{1}$  Kapteyn Astronomical Institute, University of Groningen, PO BOX 800, NL-9700 AV Groningen, the Netherlands \\
$^{2}$  INAF-Osservatorio Astronomico di Brera, via E. Bianchi 46, I-23807, Merate, Italy\\
$^{3}$  School of Physics and Astronomy, University of Southampton Highfield Campus, Southampton SO17 1PS, UK}

\date{Accepted 2020 December 18. Received 2020 December 18; in original form 2020 October 05.}

\pubyear{2020}

\begin{document}
\label{firstpage}
\pagerange{\pageref{firstpage}--\pageref{lastpage}}
\maketitle

\begin{abstract}
Spectral-timing analysis of the fast variability observed in X-rays is a powerful tool to study the physical and geometrical properties of the accretion/ejection flows in black-hole binaries. The origin of type-B quasi-periodic oscillations (QPO), predominantly observed in black-hole candidates in the soft-intermediate state, has been linked to emission arising from the relativistic jet. In this state, the X-ray spectrum is characterised by a soft-thermal blackbody-like emission due to the accretion disc, an iron emission line (in the 6--7~keV range), and a power-law like hard component due to Inverse-Compton scattering of the soft-photon source by hot electrons in a corona or the relativistic jet itself. The spectral-timing properties of \MAXI have been recently studied using observations obtained with the \NICER observatory. The data show a strong type-B QPO at $\sim$4.5~Hz with increasing fractional rms amplitude with energy and positive lags with respect to a reference band at 2--2.5~keV. We use a variable-Comptonisation model that assumes a sinusoidal coherent oscillation of the Comptonised X-ray flux and the physical parameters of the corona at the QPO frequency, to fit simultaneously the energy-dependent fractional rms amplitude and phase lags of this QPO. We show that two physically-connected Comptonisation regions can successfully explain the radiative properties of the QPO in the full 0.8--10~keV energy range. 
\end{abstract}



\section{Introduction}

Accreting Black-Hole (BH) X-ray binaries display
strong variability in a broad range of timescales, from a few tens of milliseconds to several days-months \citep[for recent reviews see][and references therein]{2016ASSL..440...61B,2020arXiv200108758I}. Different types of low-frequency quasi-periodic oscillations (QPOs) are significantly detected in the power density spectra (PDS) of these systems in the 0.1--30~Hz frequency range. These QPOs were originally classified in the BH candidate XTE J1550--564 \citep{1999ApJ...514..939W,2002ApJ...564..962R,2005ApJ...629..403C} into three main types, A, B, and C, depending on the shape of the broad-band noise component in the power spectra and the spectral state of the source. 

In particular, the type-B QPOs appear during the short-lived soft-intermediate state (SIMS), in the transition from the low-hard to the high-soft states. In this state, the energy spectrum is rather soft \citep[e.g.,][]{2016ASSL..440...61B}, dominated by a geometrically thin and optically thick accretion disc component \citep{1973A&A....24..337S} but also showing a prominent hard component possibly due to inverse-Compton scattering of the soft-photons from the disc in a corona of hot electrons \citep{1979Natur.279..506S,1997ApJ...480..735K}. 
Type-B QPOs have typical centroid frequencies in the 4--6~Hz range with fractional rms amplitudes of a few percent, and quality factors $Q \ga 6$ \citep{1999ApJ...514..939W,2005ApJ...629..403C}. 
Fast transitions can occur between different states, associated with rapid changes between the type-B QPO and other types of QPOs in the PDS \citep[e.g.,][]{1997ApJ...489..272T,2003A&A...412..235N,2005A&A...440..207B}. These transitions can be accompanied by discrete jet ejections observed in radio \citep[e.g.,][]{2004MNRAS.355.1105F}, suggesting a possible connection between the type-B QPO and the relativistic jet \citep[e.g.,][]{2009MNRAS.396.1370F, 2020ApJ...891L..29H, 2020MNRAS.495..182R}. 

Several models have been proposed to explain the phase lags in QPOs \citep{1998MNRAS.299..479L,2012ApJ...752L..25S,2013ApJ...779...71M}. Recently,
\cite{2020A&A...640L..16K} presented a quantitative explanation of the results reported by \cite{2016MNRAS.460.2796S} on the type-B QPOs in GX~339-4. In the model of \cite{2020A&A...640L..16K}, Comptonisation takes place in a jet that precesses at the QPO frequency and is fed by the hot inner flow \citep[see, e.g.,][for numerical simulations on this effect]{2018MNRAS.474L..81L}, thus producing a roughly periodic variation in the observed power-law index. A Comptonisation origin of the hard lags in the broadband noise component in the PDS of BH systems has been 
suggested by \cite{1988Natur.336..450M} for the case of the high-mass X-ray binary Cyg~X-1. On the other hand, reverberation off the accretion disc naturally leads to soft lags. Reverberation, however, cannot explain the lags of the QPO \citep{2015ApJ...814...50D}. A possible way to produce soft lags with Comptonisation is to introduce a feedback loop, where a fraction of the up-scattered photons in the corona impinge back onto the soft-photon source \citep{2001ApJ...549L.229L}. Other models \citep[e.g.,][]{2000ApJ...538L.137N,2009MNRAS.397L.101I,2016MNRAS.461.1967I} explain the phase or time lags of the QPOs but not many of these models can explain the energy-dependent rms amplitude of the variability.

\MAXI is an X-ray transient discovered on 2019 January 26 with the {\em MAXI} instrument on board the {\em ISS} \citep{2019ATel12425....1Y,2020ApJ...899L..20T} . The outburst of \MAXI has also been detected with {\em Fermi} and {\em Swift BAT} \citep{2019GCN.23795....1D,2019GCN.23796....1D}. Simultaneous observations performed with the {\em Swift XRT} led to a precise determination of the sky position and an optical counterpart was subsequently reported \citep{2019ATel12430....1D}. \NICER performed a long-term monitoring of the source, and early observations suggested that the source is a BH candidate, based on the evolution of its energy spectra and variability \citep{2019ATel12447....1S}. This was further investigated with {\em ATCA} radio data \citep{2019ATel12456....1R} and {\em INTEGRAL} hard X-ray observations \citep{2019ATel12457....1C}. Later on, a more thorough X-ray analysis of \NICER observations \citep{2020MNRAS.499..851Z} confirmed the BH-binary nature of the source. \MAXI experienced a transition to the high-soft state firstly observed with {\em MAXI} and confirmed with {\em INTEGRAL} \citep{2019ATel12471....1C}, followed by a strong radio flare found in {\em MeerKAT} observations \citep{2019ATel12497....1C}. In a recent paper, \cite{2020MNRAS.496.4366B} presented a detailed spectral-timing analysis of the type-B QPO detected in \NICER observations, in the 0.2--12~keV energy range \citep{2012SPIE.8443E..13G}, performed after the transitions. 

In this paper we use a recently-developed spectral-timing Comptonisation model \citep{2020MNRAS.492.1399K} to fit the energy-dependent fractional rms and phase lags of the type-B QPO of \MAXI. Our fitting results provide physical properties of the Comptonisation region which are not directly accessible through fits to the time-averaged spectrum. In Sect. 2 we briefly describe the observations, data analysis and measurements of the QPO and the time-averaged spectra of \MAXI published by \cite{2020MNRAS.496.4366B}. In Sec.~3 we introduce the spectral-timing Comptonisation model and our fitting methods. In Sec.~4 we present our results that we finally discuss in Sec.~5.

\section{The type-B QPO in \MAXI}

\subsection{Time-averaged spectral results}
\label{sec:time-avg}

A complete spectral analysis of a selected set of observations performed with \NICER was presented in Table~2 of \cite{2020MNRAS.496.4366B}. This data set comprises the six pointed observations of \MAXI where the type-B QPO was detected. The spectra were fitted in \XSPEC \citep{1996ASPC..101...17A} using a spectral model including Comptonisation from a disc-blackbody soft-photon source, a Gaussian to represent the iron-line emission, and an absorption component. The average temperature of the disc in these six observations was $kT_{\rm dbb} \approx 0.6$~keV with very little spread, with an area consistent with an inner radius of the disc, $R_{\rm in} \approx 250$~km \citep[for an assumed distance of 10~kpc and an assumed inclination of 60 deg, see][]{2020MNRAS.496.4366B}. The power-law index of the Comptonisation component, $\Gamma$, was approximately $3.5$. Since \NICER covers up to $\sim$10~keV, and no high-energy cutoff is seen on the data, the electron temperature in the corona could not be constrained from the available spectra ($kT_e > 10$~keV). 

\subsection{Spectral-Timing of the type-B QPO}

The energy-dependent fractional-rms amplitude and phase lags of the type-B QPO of \MAXI obtained from \NICER observations were presented in Figure~4 of \cite{2020MNRAS.496.4366B}. The QPO is detected in the full energy range, from 0.75 to 10~keV, with increasing fractional rms amplitude, going from $\la$1\% below 2~keV to 10--12\% at the highest energy channels. Taking the 2--2.5~keV band as reference band, the lag-energy spectrum shows positive lags for every energy channel. This means that photons with energies lower and higher than the reference band lag behind the photons in the 2--2.5~keV band. The softest photons have a lag of $\sim$0.9~rad, while the hardest ones have a lag of $\sim$0.6~rad. At a QPO frequency $\nu_0 = 4.45$~Hz, those phase lags correspond to time lags between $\sim$32~ms and $\sim$21~ms, respectively.

\section{The spectral-timing Comptonisation model}
\label{sec:model}

In this work, we use the spectral-timing Comptonisation model from \cite{2020MNRAS.492.1399K} and \paperkostasalt, which is based on the idea proposed by \cite{1998MNRAS.299..479L}, \cite{2001ApJ...549L.229L}, and \cite{2014MNRAS.445.2818K}. This family of models are built to explain the radiative properties of any oscillation of the time-averaged or steady-state spectrum, $SSS(E)$, which arises due to the coupled oscillations of the physical parameters of the system, such as the temperature of the corona and that of the source of seed photons.
During the inverse-Compton process, the seed photons gain energy from the electrons in the corona, that hence have to cool down. However, since the corona of these astrophysical sources are long-lived, a heating source, the so-called external heating rate, $H_{\rm ext}$, has to operate  leading to a thermal balance \citep{2006MNRAS.370..405S}. In this model, the origin of the QPO frequency is not explained, but it is assumed to be a sinusoidal coherent oscillation of the Comptonised X-ray flux and the physical parameters of the corona at the QPO frequency, $\nu_0$. The strength of the model lies in its ability to describe the energy-dependent spectral-timing properties of the QPO, given by the rms amplitudes and phase lags which, in turn, can be used to estimate the physical properties of the corona.

The model of \cite{2020MNRAS.492.1399K} assumes that the source of seed photons is a spherical blackbody of temperature, $kT_s$, enshrouded by a spherically-symmetric and homogeneous Comptonising region (the corona), with a characteristic size, $L$, temperature, $kT_e$, and optical depth, $\tau$. The model also incorporates the feedback of the up-scattered photons onto the soft-photon source, parametrized by a so-called feedback fraction, $0 \le \eta \le 1$, defined as the fraction of the flux of the seed-photon source due to the feedback process. The steady state (time-averaged) spectrum is calculated by solving the stationary Kompaneets equation \citep{1957JETP....4..730K}. In order to obtain the variability amplitude and phase of the QPO, the model solves the linearised time-dependent Kompaneets equation, assuming that $kT_e$, $kT_s$, and $H_{\rm ext}$ undergo oscillations at the QPO frequency. This yields the complex variability amplitude of the spectrum, $N(E) = |N(E)| e^{i \phi(E)}$, where $|N(E)|$ is the absolute rms amplitude and $\phi(E)$ is the phase lag. The fractional rms amplitude can be calculated normalising the absolute rms amplitude by the flux of the  steady-state solution, $n(E) = |N(E)| / SSS(E)$.

\cite{2020MNRAS.492.1399K} proved that the model can fit the energy-dependent time lags and fractional rms amplitude of the kilohertz QPOs in the neutron-star low-mass X-ray binary (LMXB) 4U~1636--53. Moreover, \paperkostasalt showed that the model is also suitable to explain the type-C QPO in the BH LMXB GRS~1915+105. In this paper we apply the same model to fit the spectral-timing properties of the type-B QPO observed in \MAXI data. The model assumptions include (see the next Sections for a more detailed discussion) a blackbody-like seed-photon source and a spherically symmetric and homogeneous corona, which are rough approximations to an accretion disc, a real corona and a jet, all of which are highly asymmetrical components. The properties inferred from the fits can then be taken as characteristic quantities that not only allow us to understand the system in more detail, but also serve as a guide of the necessary steps to build a more realistic and complete model.

In order to fit the model to the data, we use a Markov-Chain Monte Carlo (MCMC) scheme based on the affine-invariant ensemble sampler \emcee \citep{2013PASP..125..306F}. With this approach we can fit the fractional rms and phase-lag energy spectra, either independently or simultaneously, starting from a broadly-spread set of walkers or from a group of walkers distributed around a certain maximum of the posteriors. In the rest of the paper we will assume that the seed-photon source is a sphere with a radius of 250~km, consistent with the inner radius of the disc blackbody fitted to the time-averaged spectra \citep{2020MNRAS.496.4366B}.

\section{Results}
\label{sec:results}

\begin{figure}
  \includegraphics[angle=000,width=\columnwidth]{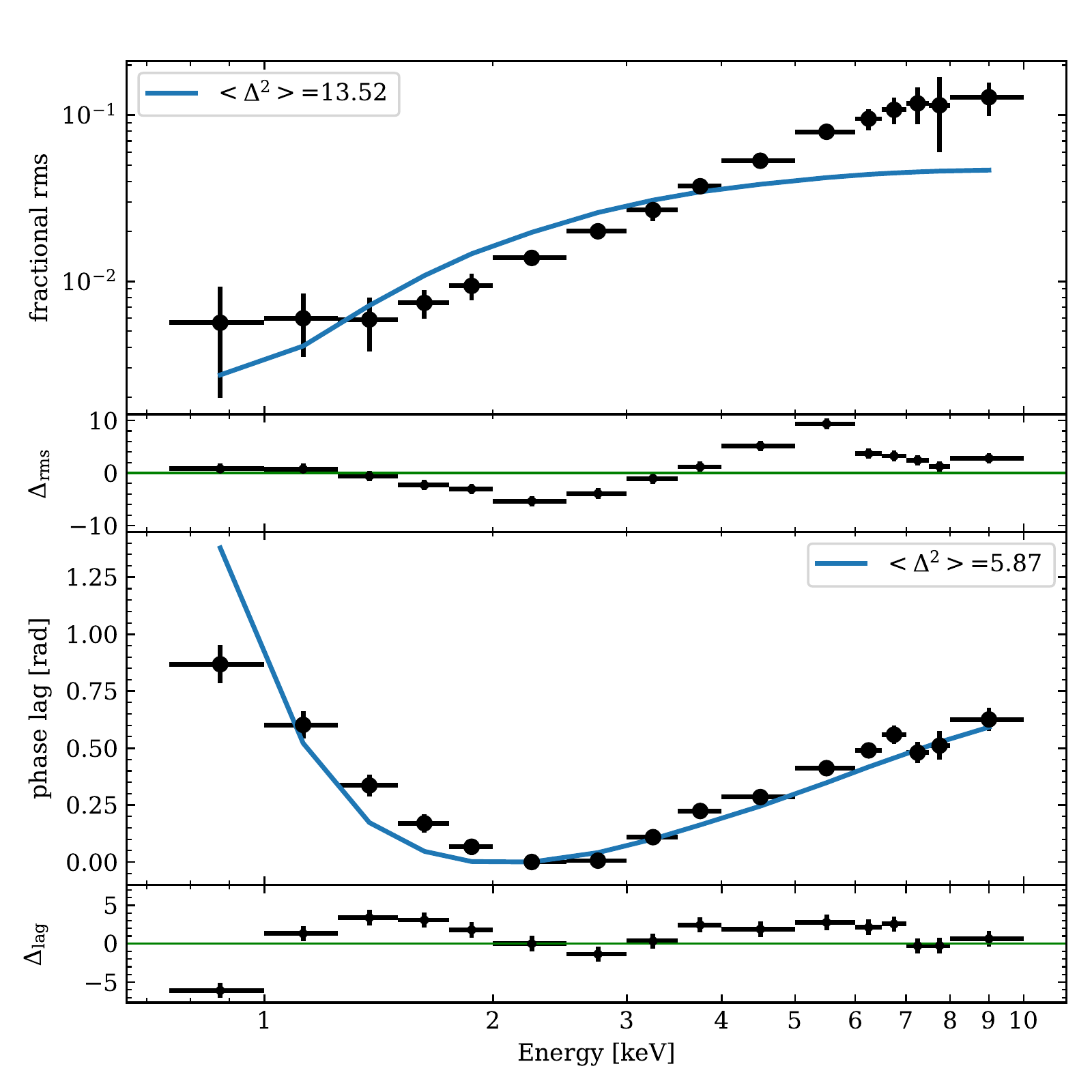}
  \caption{Fractional rms (upper panel) and phase-lag spectra (lower panels) of the 4.45~Hz type-B QPO of \MAXI. The solid lines represent the best-fitting model obtained for a corona size of $\sim$7\,100~km and a $\sim$53\% feedback fraction onto a soft-photon source with a temperature of $\sim$0.2~keV. Residuals, $\Delta$~=~(data--model)/error, are also shown. Average quadratic residuals, $\Delta^2$, are indicated at the top of each panel.}
 \label{fig:fitting}
\end{figure}

\begin{figure*}
  \includegraphics[angle=000,width=1.25\columnwidth]{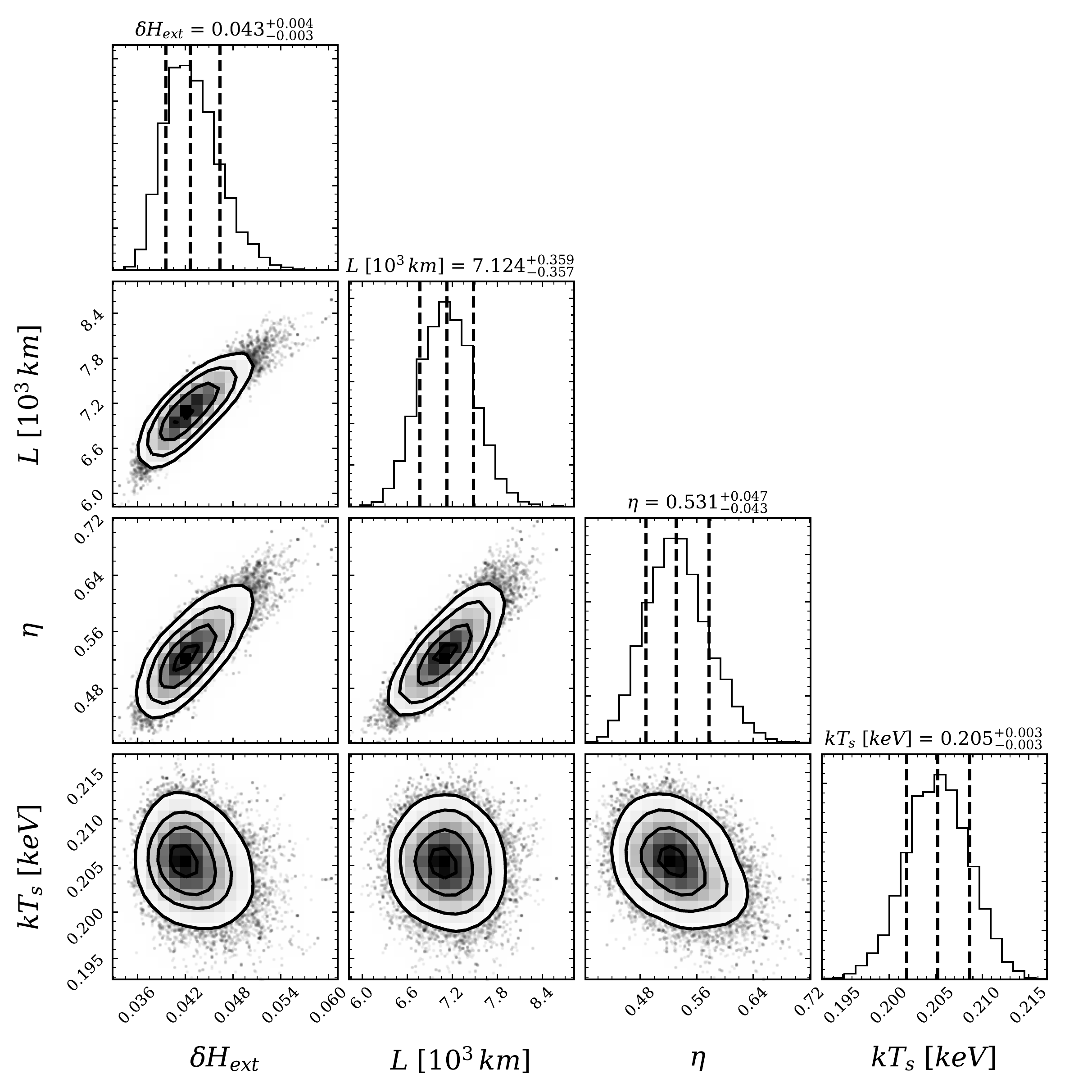}
  \caption{Corner plot of the physical parameters of \MAXI obtained with a single Comptonisation model, including the amplitude of the external heating rate, $\delta H_{\rm ext}$, the size, $L$, the temperature of the soft-photon source, $kT_s$, and the feedback fraction, $\eta$.}
 \label{fig:corner}
\end{figure*}

\begin{table*}
 \caption{Best-fitting values of the physical parameters of \MAXI and their corresponding 1-$\sigma$ (68\%) uncertainties obtained with a single-component corona.}
 \label{tab:table}
 \begin{tabular}{ ccc | ccccccc }
  \hline
  $kT_e$ & $\Gamma$ & $\tau$ & $kT_s$ & $L$ & $\eta$ & ${\delta H}_{\rm ext}$ & $\delta kT_s$ & $\delta kT_e$ & $\chi_\nu^2$ (dof)\\
  (keV) & & & (keV) & (km) & & (\%) & (\%) & (\%)   \\
  \hline
  20$^\dagger$ & 3.5$^\dagger$ & 1.3$^\dagger$ & 0.205$\pm$0.003 & 7\,100$\pm$360 & 0.53$\pm$0.05 & 4.3$\pm$0.4 & 0.12$\pm$0.01 & 1.89$\pm$0.09 & 11.1 (28 dof) \\
  30$^\dagger$ & 3.5$^\dagger$ & 0.94$^\dagger$ & 0.202$\pm$0.004 & 7\,700$\pm$370 & 0.50$\pm$0.04 & 5.2$\pm$0.6 & 0.12$\pm$0.01 & 1.53$\pm$0.07 & 11.2 (28 dof) \\
  50$^\dagger$ & 3.5$^\dagger$ & 0.62$^\dagger$ & 0.183$\pm$0.004 & 8\,100$\pm$400 & 0.42$\pm$0.04 & 10.3$\pm$1.5 & 0.14$\pm$0.01 & 1.11$\pm$0.05 & 12.1 (28 dof) \\
  \hline
  $^\dagger$ fixed parameters.
 \end{tabular}
\end{table*}

In Figure~\ref{fig:fitting} we present the best-fitting model found from the simultaneous fit of the rms (top panel) and lag (bottom panel) spectra. Despite the simplicity of the underlying assumptions (see Sec.~\ref{sec:model}), the model recovers the main features of the spectrum notably: i) the rms amplitude in the model increases with energy, spanning from a non-vanishing rms ($\la$0.5\%) at low energies to a maximum rms amplitude of 4--5\% at the highest energy channels; ii) the lags with respect to the reference band (2--2.5~keV) are all positive, with a shape of the lag spectrum that is very similar to the observed one. However, the residuals found are very significant, yielding a reduced $\chi_\nu^2 = 11.1$ for 28 degrees of freedom (dof). At the top of each panel of Fig.~\ref{fig:fitting} we give the average quadratic residuals of the fits, $\left<\Delta^2\right> = 1/N \sum_{i=1}^N (x-x_i)^2/\sigma_i^2$, where $x$ is the model, and $x_i$ and $\sigma_i$ are the data and their corresponding 1-$\sigma$ error bars. Since the relative errors of the rms are significantly smaller than those of the lags, we arbitrarily increased the error-bars of the rms spectrum by a factor of 5 to make the fitting process more sensitive to the lag spectrum. The best-fitting model was obtained fixing the Comptonisation power-law index to $\Gamma=3.5$, and the temperature of the corona to $kT_e=20$~keV, guided by the values obtained from the time-averaged spectra (see Sec.~\ref{sec:time-avg}). Those values set the optical depth to 
$\tau = \sqrt{ 2.25 + \frac{3}{\frac{kT_e}{m_e c^2}\left[(\Gamma+0.5)^2-2.25\right]}} = 1.3$. Here $m_e$ and $c$ are, respectively, the rest mass of the electron and the speed of light.
The corona size and feedback fraction best-fitting values, $L=7\,100\pm360$~km and $\eta=0.53\pm0.05$, respectively, are mainly set by the lag spectrum, while the temperature of the soft-photon source, $kT_s = 0.205\pm0.003$~keV, and the amplitude of the variability of the external-heating rate, $\delta H_{\rm ext}=4.3\pm0.4$\%, are driven by the rms spectrum. For this best-fitting solution, the fractional amplitudes of the temperatures were $\delta kT_s = 0.12\pm0.01$\% and $\delta kT_e = 1.9\pm0.1$\%, respectively. 
We also fitted the data fixing $kT_e=30$~and~50~keV, with their corresponding $\tau$ values. We found very similar solutions in the 0.8--10~keV range (with similar $\chi^2$ values), and best-fitting physical parameters fully consistent with those found using $kT_e=20$~keV, with the differences in the best-fitting values (less than 1--2$\sigma$) due mainly to the change in the optical depth, $\tau$, due to the different values of $kT_e$ used with the same $\Gamma$ (see above).
We also tried to fit the data by letting the values of $\Gamma$ and $kT_e$ free, but the fitting did not improve significantly, as expected, given the relatively low upper-energy bound of the \NICER instrument. In Figure~\ref{fig:corner} we present the corner plot resulting from the MCMC chain analysis. As usual, the diagonal in this plot shows the posterior probabilities of the physical parameters of the model with the median value and the 1-$\sigma$ confidence levels indicated in the plots. The lower triangle in this figure shows the joint probabilities of each pair of parameters with their corresponding contours at 68, 90 and 95\% levels. We summarise the results for the best-fitting parameters and their corresponding uncertainties in Table~\ref{tab:table}, including the three values explored for $kT_e$.


\subsection{A two-components `\dual' model}
\label{sec:dualmodel}

\begin{figure*}
  \includegraphics[angle=000,width=0.99\columnwidth]{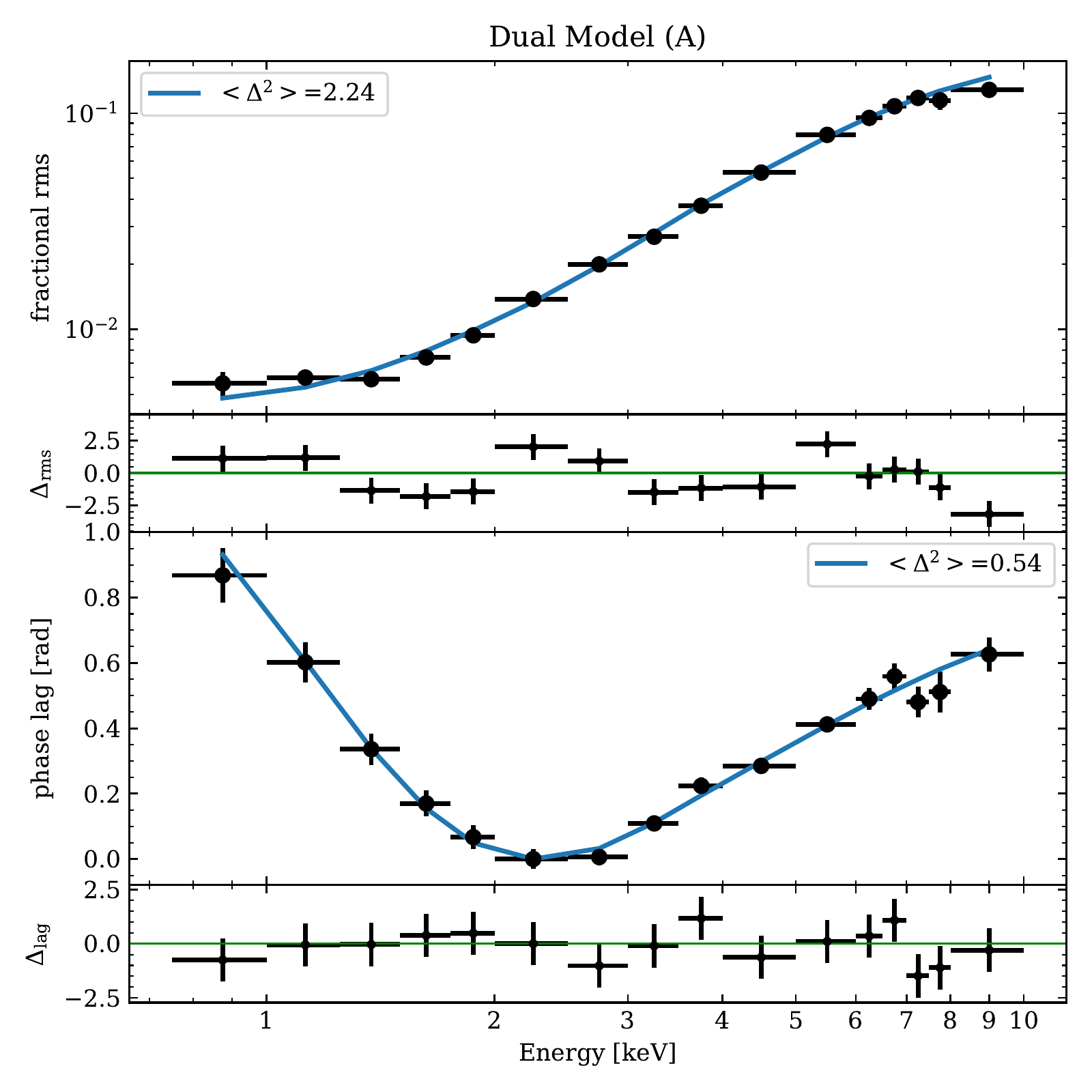}
  \includegraphics[angle=000,width=0.99\columnwidth]{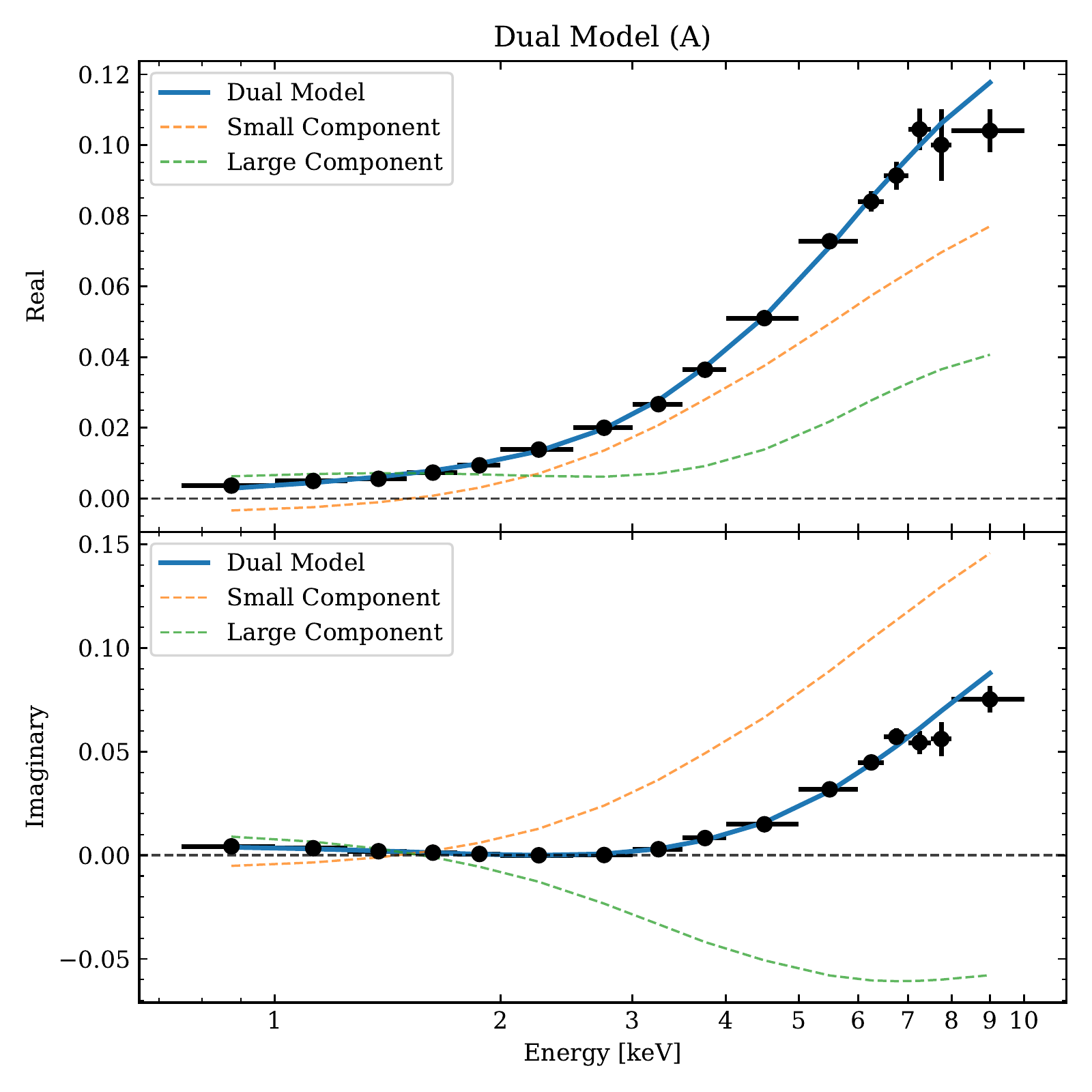}
  \caption{Energy-dependent fractional rms amplitude (upper panels) and phase lags (lower panels) of the 4.45~Hz type-B QPO of \MAXI. The best-fitting models (solid lines) are obtained using a combination of two Comptonisation components (\dual model, Case A). Residuals, $\Delta$~=~(data--model)/error, are also shown. Average quadratic residuals, $\Delta^2$, are indicated at the top of each panel.  {\bf Right panel:} Energy-dependent Real (top panel) and Imaginary (bottom panel) parts of the complex oscillation. The individual contribution of the {\em small} ({\em large}) component is plotted with dashed lines in orange (green) colour.}
 \label{fig:fitting_dualA}
\end{figure*}

\begin{figure*}
  \includegraphics[angle=000,width=1.99\columnwidth]{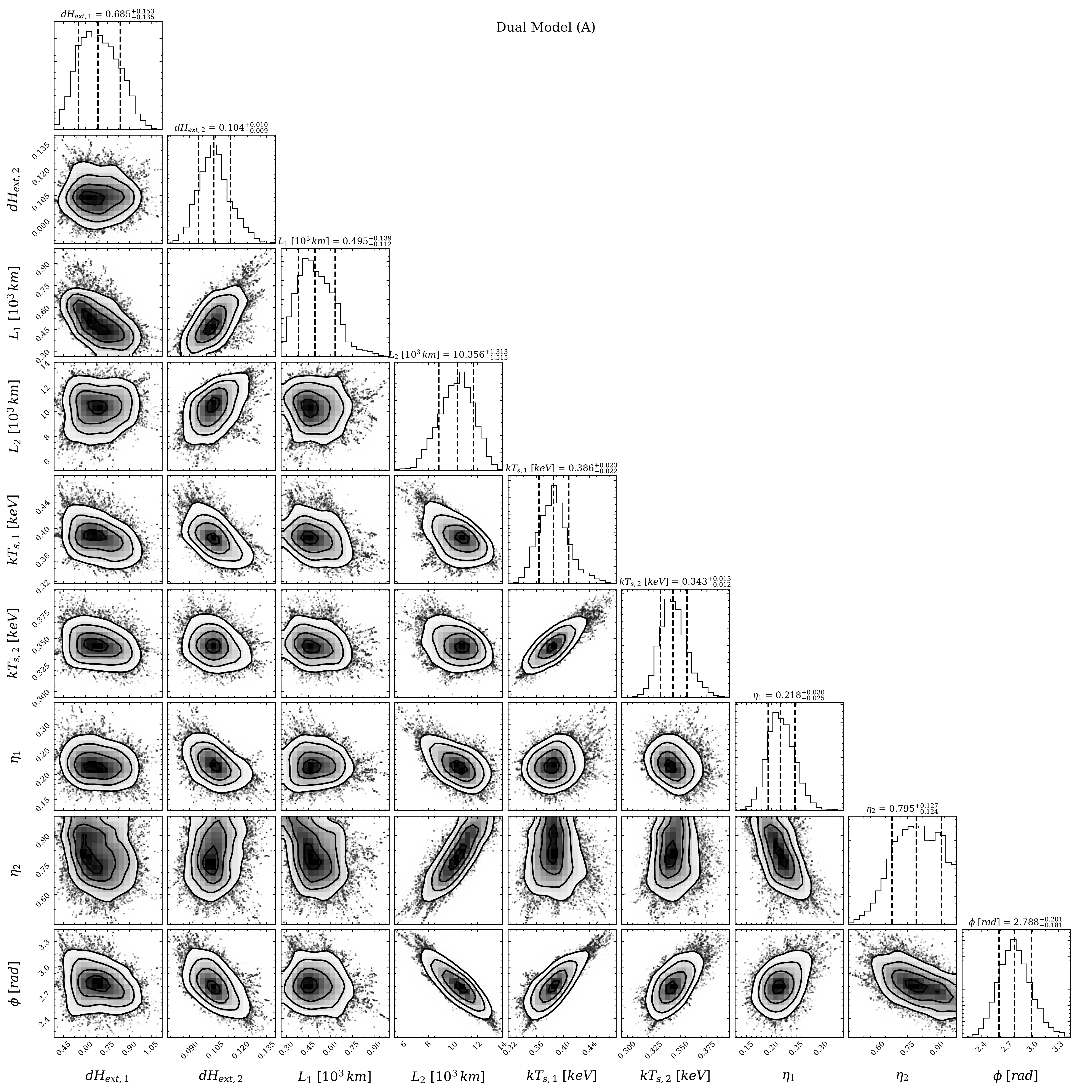}
  \caption{Corner plot of the physical parameters of \MAXI obtained with the MCMC scheme using the \dual model for the Case A. The model involves a {\em small (1)} and a {\em large (2)} Comptonisation region, characterised by the amplitudes of external heating rates, $\delta H_{{\rm ext}, 1-2}$, sizes, $L_{1-2}$, temperatures of the soft-photon sources, $kT_{s,1-2}$, and feedback fractions, $\eta_{1-2}$), respectively, and a phase angle, $\phi$, between both oscillating components.}
 \label{fig:corner_dualA}
\end{figure*}

\begin{table*}
 \caption{Best-fitting values for the physical parameters of \MAXI and their corresponding 1-$\sigma$ (68\%) uncertainties of the \dual model (Case A).}
 \label{tab:table_dualA}
 \begin{tabular}{ l | ccc | cccccc }
  \hline
  Component & $kT_e$ & $\Gamma$ & $\tau$ & $kT_s$ & $L$ & $\eta$ & ${\delta H}_{\rm ext}$ & $\delta kT_s$ & $\delta kT_e$\\
  & (keV) & & & (keV) & ($10^3$~km) & & (\%) & (\%) & (\%) \\
  \hline
  {\em Case A:} ~~ $\chi^2_\nu$ = 1.92 (23 dof) ~~ $\phi$ = 2.79$\pm$0.20 rad\\
  Small (1) & 20$^\dagger$ & 3.5$^\dagger$ & 1.3$^\dagger$ & 0.38$\pm$0.02 & 0.49$\pm$0.14 & 0.22$\pm$0.03 & 68$\pm$15 & 1.1$\pm$0.3 & 42$\pm$10 \\
  Large (2) & 20$^\dagger$ & 3.5$^\dagger$ & 1.3$^\dagger$ & 0.34$\pm$0.01 & 10.4$\pm$1.5 & 0.80$\pm$0.12 & 10$\pm$1 & 0.42$\pm$0.05 & 8.8$\pm$0.8  \\
  \hline
  
  {\em Case A:} ~~ $\chi^2_\nu$ = 1.96 (23 dof) ~~ $\phi$ = 2.83$\pm$0.24 rad \\
  Small (1) & 30$^\dagger$ & 3.5$^\dagger$ & 0.94$^\dagger$ & 0.39$\pm$0.03 & 0.48$\pm$0.11 & 0.22$\pm$0.03 & 70$\pm$20 & 1.3$\pm$0.3 & 48$\pm$10\\
  Large (2) & 30$^\dagger$ & 3.5$^\dagger$ & 0.94$^\dagger$ & 0.35$\pm$0.01 & 10.5$\pm$1.6 & 0.76$\pm$0.14 & 10$\pm$1 & 0.42$\pm$0.05 & 7.5$\pm$0.8\\
  \hline
  {\em Case A:} ~~ $\chi^2_\nu$ = 2.12 (23 dof) ~~ $\phi$ = 2.77$\pm$0.25 rad \\
  Small (1) & 50$^\dagger$ & 3.5$^\dagger$ & 0.62$^\dagger$ & 0.40$\pm$0.03 & 0.45$\pm$0.12 & 0.18$\pm$0.03 & 78$\pm$18 & 1.2$\pm$0.4 & 50$\pm$13\\
  Large (2) & 50$^\dagger$ & 3.5$^\dagger$ & 0.62$^\dagger$ & 0.35$\pm$0.01 & 11.8$\pm$1.7 & 0.76$\pm$0.12 & 10$\pm$1 & 0.41$\pm$0.05 & 6.0$\pm$0.8\\
  \hline
  $^\dagger$ fixed parameters.
 \end{tabular}
\end{table*}

Given the large $\chi^2$ value that we obtained by fitting the variability spectrum of the QPO with a single-component Comptonisation model, and considering the simplified assumptions of the model (e.g., a spherical corona with uniform density), we decided to explore the possibility that the variability spectra of the type-B QPO arise from Comptonisation occurring in two different, but physically connected, regions, namely a {\em small (1)} and a {\em large (2)} region located in the vicinity of the compact object. We call this the `\dual' model.

In order to obtain the fractional rms amplitude and phase lags of the combination of two oscillating models as a function of energy, we sum the contributions of each component in the complex Fourier space in absolute units and re-normalise to fractional units using the count-rate at each energy band given by the addition of the two steady-state spectra predicted by the model, $SSS_1(E)$ and $SSS_2(E)$. This is possible thanks to the linearity of the Fourier Transform (not the power spectrum). The linear combination of two sinusoidal functions (or complex exponentials) at the same frequency is a new sinusoidal with amplitude and phase determined by the relative amplitudes and phases of the two underlying sinusoidal functions. In the complex plane, this can be expressed as:

\begin{align} \label{eq:oscillation}
    N(E, t, \omega_0) = &\, N_1(E) e^{i \omega_0 t} + N_2(E) e^{i (\omega_0 t + \phi)} \\
    = & \left(N_1(E) + N_2(E) e^{i \phi}\right) e^{i \omega_0 t} = |N(E)| e^{i \Phi(E)} e^{i \omega_0 t}, \nonumber
\end{align}

\noindent where $N_1(E)$ and $N_2(E)$ are the complex energy-dependent amplitudes of the oscillation at the QPO frequency, $\omega_0 = 2\pi\nu_0$, in absolute-rms units, delayed by a fixed phase angle, $\phi$. $N(E, t)$ is the arising complex energy-dependent light curve which, as shown, is another sinusoidal function with amplitude $|N(E)|$ and phase angle $\Phi(E)$. If the underlying complex amplitudes are expressed in polar form,  $N_1(E) = |N_1(E)| e^{i \phi_1(E)}$ and $N_2(E) = |N_2(E)| e^{i \phi_2(E)}$, respectively, the resulting amplitude and phase of $N(E, t)$ are given by:

\begin{multline} \label{eq:dual_rms}
|N(E)| = \big[ |N_1(E)|^2 + |N_2(E)|^2 \\
      - 2 |N_1(E)| |N_2(E)| \cos(\phi_2(E) - \phi_1(E) + \phi) \big]^{1/2} , 
\end{multline}

\begin{equation} \label{eq:dual_lag}
\tan \left(\Phi(E)\right) = \frac{{\rm Im} \lbrace N_1(E) \rbrace + {\rm Im} \lbrace N_2(E) e^{i\phi} \rbrace}{ {\rm Re} \lbrace N_1(E) \rbrace + {\rm Re} \lbrace N_2(E) e^{i\phi}\rbrace} ,
\end{equation}

\noindent where Re and Im are the real and imaginary parts of the complex amplitudes, respectively. Finally, we divide the absolute rms amplitudes by the full steady-state solution to obtain the fractional-rms amplitude: $n(E)=|N(E)|/(SSS_1(E)+SSS_2(E))$. Hence, the resulting rms and lag spectra of the combination of two Comptonisation models can be obtained from the underlying components by incorporating a new free parameter: the delay among them, defined as a the phase angle $\phi$ between the two models at the reference band. Once this angle is known, the full set of subject bands in the variability spectra are uniquely defined.

After we set up these equations into our MCMC framework, we ran a long chain of 2\,000 steps for 240 uniformly distributed walkers to perform a wide exploration of the full parameter space, comprising $\delta H_{\rm ext, 1}$, $kT_{s, 1}$, $L_1$, $\eta_1$ and $\delta H_{\rm ext, 2}$, $kT_{s, 2}$, $L_2$, $\eta_2$ for the {\em small} and {\em large} regions, respectively, and the phase angle $\phi$ between each other. As in the single-component model, in these runs we fixed $kT_e = 20$~keV and $\Gamma = 3.5$ to keep consistency with the time-averaged spectra. For these fits we did not increase the error bars of the fractional rms spectrum. This parameter-space exploration led to two clusters of solutions in different regimes: Case ``A'' ($\phi \sim 2.8$~rad) and Case ``B'' ($\phi \sim 0.2$~rad). In order to explore in detail the two clusters of solutions, we ran two MCMC explorations using a set of 240 walkers initially located around the best-fitting solutions found for each cluster, and we let them walk for 2\,000 steps each. 

The \dual models fit both the rms and lag spectra significantly better than the previous model, 
with reduced $\chi_\nu^2 = 1.92$ and 1.32, respectively, for 23~dof. Despite the fact that a better fit can be naturally expected because of the extra free parameters in the \dual model, the results are remarkably better and now the full rms and lag spectra can be fitted with great accuracy 
(see Fig.~\ref{fig:fitting_dualA}). In Case A, the {\em small} component, $L_1 = 490\pm14$~km, has a low feedback fraction, $\eta_1 = 0.22\pm0.03$, while the {\em large} component, $L_2 = 10.4\pm1.5 \times 10^3$~km, requires a high feedback, $\eta_2 = 0.80\pm0.12$, to explain the observed lags. The best-fitting soft-photon source temperatures are $kT_{s, 1} = 0.38\pm0.02$ and $kT_{s, 2} = 0.34\pm0.01$~keV, consistent with each other, indicating a possible common origin of the soft photons that are subsequently Comptonised. These blackbody temperatures are consistent with higher disc-blackbody temperatures, which together with the adopted values of $\Gamma$ and $kT_e$, make the steady-state spectrum arising from our variable-Comptonisation model fully consistent with the actual time-averaged spectrum fitted by \cite{2020MNRAS.496.4366B}. We also fitted the data fixing $kT_e$ to 30 and 50~keV, as in the single-component case. We found similar solutions with consistent $\chi^2$ values and best-fitting parameters with respect to those found for $kT_e=20$~keV. The fractional amplitudes of the temperatures of the {\em small} component, $\delta kT_{s, 1} \sim 1.2$\% and $\delta kT_{e, 1} \sim 45$\%, and of the external heating rate, $\delta H_{\rm ext, 1} \sim 68$\%, are quite high, but still constrained. In turn, the fractional amplitudes of the {\em large} component are small, all below $\lesssim 10$\%.

For Case B, the Comptonising regions are described by a {\em small} component with size $L_1 = 2\,200\pm240$~km and a high feedback fraction, $\eta_1 = 0.89\pm0.04$, and a {\em large} region with size $L_2 = 21\pm3 \times 10^3$~km, and a low feedback fraction, $\eta_2 < 0.22$, consistent with zero. The best-fitting soft-photon source temperatures $kT_{s, 1} = 0.48\pm0.03$ and $kT_{s, 2} = 0.73\pm0.02$~keV, are slightly higher, possibly indicating a bigger impact of the hotter inner regions of the accretion disc, given the temperature values inferred from the time-averaged spectrum. 
In this case, the fractional amplitudes of the temperatures of the {\em small} component, $\delta kT_{s, 1} \sim 74$\% and $\delta kT_{e, 2} \sim 355$\%, and of the external heating rate, $\delta H_{\rm ext, 1} \sim 550$\%, are very large, and thus, even if they represent mathematical solutions compatible with the observables, they break down the perturbative assumptions of the model, and thus we do not consider them further. For completeness, we present this solution in Appendix~\ref{sec:appendix}.

In Figure~\ref{fig:fitting_dualA} we present the two-component best-fitting {\em dual} model (Case A). On the left panels we show the energy-dependent fractional rms (upper panel) and phase lags (lower panel). Weighted-average residuals are indicated at the top of each panel. On the right panels we plot the Real (upper panel) and Imaginary (bottom panel) parts of the complex oscillation. On these panels we also include the contribution of each individual component, using coloured dashed lines. The {\em large} component dominates the variability below $\sim$2~keV, whereas above that energy the {\em small} component prevails. The {\em small} component has a pivot point around an energy of $\sim$1.5~keV, which reflects in a change of sign of both the real and imaginary parts of the oscillation, and a minimum of the rms amplitude. In Figure~\ref{fig:corner_dualA} we show the corner plot obtained for the MCMC chain of walkers concentrated around the best-fitting {\em dual} model (Case A). The best-fitting results and their corresponding uncertainties for each solution are presented in Table~\ref{tab:table_dualA} for the three explored values of $kT_e$.

\section{Discussion}
\label{sec:discussion}

We have shown that the spectral-timing radiative properties of the type-B QPO of \MAXI measured with \NICER can be explained by Comptonisation. We use a recently-developed variable-Comptonisation model \citep{2020MNRAS.492.1399K} to fit the energy-dependent fractional rms amplitude and phase-lag spectra of the QPO in the full 0.8--10~keV energy range. Considering one Comptonisation region, we obtain a good fit of the lag spectrum and a not so good fit of the rms spectrum, with rms-amplitudes in the hard energy bands that are lower than the values observed. If we consider two independent, but physically coupled, Comptonisation regions, we achieve excellent fits both to the rms-amplitude and phase-lag spectra. 
This is the first time that a self-consistent spectral-timing model of Comptonisation is successfully applied to data at energies below $\sim$2 keV. 

The idea that Comptonisation can explain the radiative properties of QPOs is born on the evidence that the rms amplitude of different types of QPOs in different sources increases with energy \citep[e.g.,][and references therein]{2006MNRAS.370..405S,2013MNRAS.435.2132M}. Time-dependent Comptonisation models were originally developed to explain the rms and lag spectra of kHz QPOs in neutron-star LMXBs \citep{1998MNRAS.299..479L, 2001ApJ...549L.229L, 2014MNRAS.445.2818K}. Recently, \cite{2020MNRAS.492.1399K} presented an updated version of a time-dependent Comptonisation model with many computational advances, which allows to fit efficiently the spectral-timing data of QPO observations of very different sources. 
Our best-fitting Comptonisation model (Fig.~\ref{fig:fitting}) yields a corona of $\sim$7\,000~km with a moderate feedback fraction, $\eta \approx 0.5$, and a rather low soft-photon source temperature, $kT_s \approx 0.2$~keV, not fully consistent with the temperature of the best-fitting disc blackbody to the time-averaged spectra ($kT_{\rm dbb} \approx 0.6$~keV). Despite the fact that the model reproduces the lag spectrum reasonably well, it fails to reproduce the rms spectrum accurately, leading to $\chi^2_\nu \sim 11$. This is somehow expected given the simplicity of the model and the assumptions made for its calculation (i.e. homogeneous corona, fixed optical depth, etc).

Subsequently, we fitted the variability spectrum at the QPO frequency as the combined effect of two independent, but physically coupled, Comptonisation regions. \citep[For a different model involving two Comptonisation regions see][]{2000ApJ...538L.137N}. By combining the output of two Comptonisation models in the complex plane (Fourier space of the signal, limited to the QPO frequency), we showed how to compute the energy-dependent fractional-rms and phase-lag spectra that can be compared to the data. 
The fitting results found are significantly better: both the full rms and lag spectra can be modelled with great accuracy with two sets of physical parameters, that we call cases A and B, yielding $\chi^2_\nu \approx 1.9$~and~1.3, respectively. 

In case A, the best-fitting time-dependent Comptonising model consists of a {\em small} region, $L_1 \approx 500$~km, with rather low feedback, $\eta_1 \approx 0.2$, and a {\em large} region, $L_2 \approx 10\,000$~km, with high feedback, $\eta_2 = 0.80\pm0.12$. Moreover, despite that both soft-photon source temperatures were let free during the fit, they converged to essentially equal values, $kT_{s, 1-2} \sim 0.35$~keV, pointing to a common source of soft-photons leading the full oscillation, associated to the accretion disc. These temperatures are lower than the temperature fitted to the disc blackbody component in the time-averaged spectra, 
but given the difference in the underlying spectral model we used (blackbody instead of disc blackbody), our results turn out to be compatible with those in \cite{2020MNRAS.496.4366B}, also evidencing the need of a more suitable variable-Comptonisation model including a disc-blackbody like spectrum as the soft-photon source. These best-fitting results, together with the fact that in the fit we have fixed $\Gamma = 3.5$ and $kT_e = 20$~keV, make the steady-state spectrum associated to our variability model, also consistent with the observed time-averaged spectra of \MAXI. Moreover, as shown in Sec.~\ref{sec:dualmodel}, these results are not very sensitive to changes in the assumed corona temperature. The best-fitting physical parameters found fixing $kT_e = 30$ and 50~keV remain consistent with those found assuming $kT_e = 20$~keV.

The low feedback, $\eta_1 \approx 0.2$, of the {\em small} corona component is in accordance with its relatively small size, given that $L_1 \sim 30\,R_g$ (where $R_g$ is the gravitational radius of a 10~M$_\odot$ BH) is comparable with the inner radius of the accretion disc (see Sec.~\ref{sec:time-avg}). In this configuration only a relatively small fraction of the Comptonised photons can efficiently impinge back onto the soft-photon source. On the contrary, the {\em large} region of the corona ($L_2 \sim 700\,R_g$) enshrouds the full soft-photon source and consequently it establishes a strong feedback loop ($\eta_2 \approx 0.8$). The model assumes that the corona is homogeneous and that its temperature oscillates coherently at the QPO frequency. This might possibly be hard to achieve for such a large component. Regarding the amplitudes of the external heating rate that balance the internal energy of the electrons, $\delta H_{\rm ext}$, since the {\em small} region is on average closer than the {\em large} region to the soft-photon source that is leading the variability, the ratio of $\delta H_{\rm ext, 1}/\delta H_{\rm ext, 2} \sim 6$ is naturally expected in such a scenario.

It is important to note that our model lacks a feedback interaction between the two Comptonisation regions. The oscillations of both components are related by a phase delay which, in essence, is connected to their relative delays with respect to the leading oscillation in the soft-photon source. A two-component corona was also proposed by \cite{2000ApJ...538L.137N} to explain the lags seen in the type-C QPO in the BH LMXB GRS~1915+105. In their model, hard lags are produced by Compton up-scattering of disc photons in an inner optically thick corona component. Those photons later Compton down-scatter in an outer cooler and optically thinner corona, causing soft lags. The model fits the lags of the QPO in that source but does not explain the fractional rms of the QPO. 

Despite all the limitations and caveats, we have shown that by incorporating the interaction between two different physical components, the Comptonisation model can successfully explain both the time-averaged spectrum of \MAXI and the spectral-timing properties of the type-B QPO. This is of course a simplification of a possibly more complex picture, which could be represented by a unique component with continuous extended properties, e.g., a disc-like soft-photon source, an extended corona with temperature and density gradients, and possibly angle-dependent optical depths. Our results serve as a guide for future developments of Comptonisation models. Our model could be tested with spectral-timing data of the Type-B QPO in a wider energy range, for instance, using simultaneous data from \NICER and {\em HXMT}, which covers up to $\gtrsim$100~keV \citep{2020NatAs.tmp..184M}. In this sense, this paper is a step towards the goal of building physically-motivated models capable to fit simultaneously the time-averaged and variability spectra of LMXBs at different frequencies. These models will be useful to interpret the observations of the upcoming generation of large effective area and high-time resolution X-ray instruments like {\em eXTP} \citep{2016SPIE.9905E..1QZ} and {\em Athena} \citep{2013arXiv1306.2307N}.

\section*{Acknowledgements}

We acknowledge the Referee for constructive comments, which helped us to improve our manuscript. This work is part of the research programme Athena with project number 184.034.002, which is (partly) financed by the Dutch Research Council (NWO). TMB acknowledges financial contribution from the agreement ASI-INAF n.2017-14-H.0. LZ acknowledges support from the Royal Society Newton Funds. DA acknowledges support from the Royal Society.

\section*{Data Availability}

This research has made use of data obtained from the High Energy Astrophysics Science Archive Research Center (HEASARC), provided by NASA’s Goddard Space Flight Center.

\bibliographystyle{mnras}
\bibliography{example}

\appendix

\section{A secondary solution for the two-components {\em dual} model} \label{sec:appendix}

\begin{figure*}
  \includegraphics[angle=000,width=0.99\columnwidth]{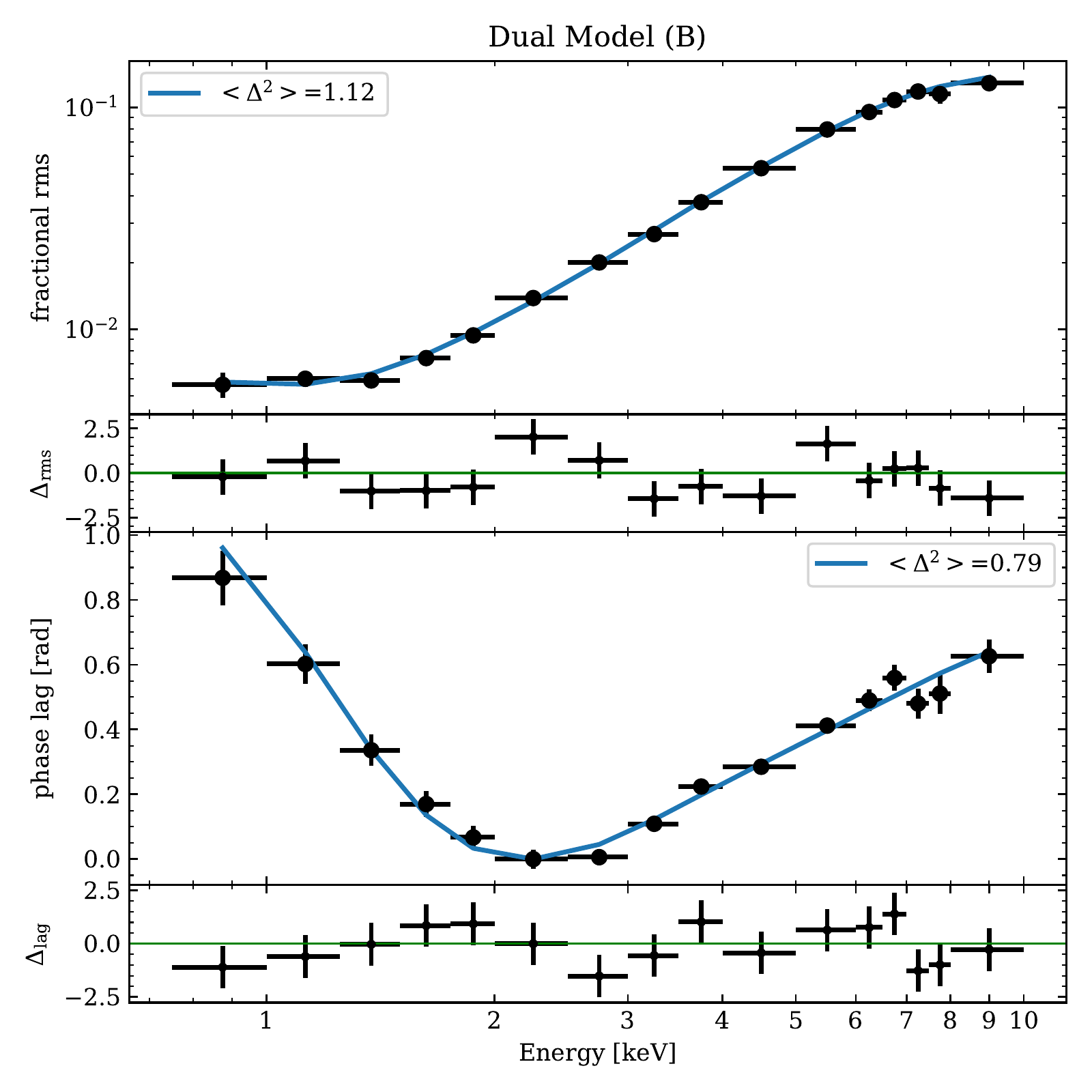}
  \includegraphics[angle=000,width=0.99\columnwidth]{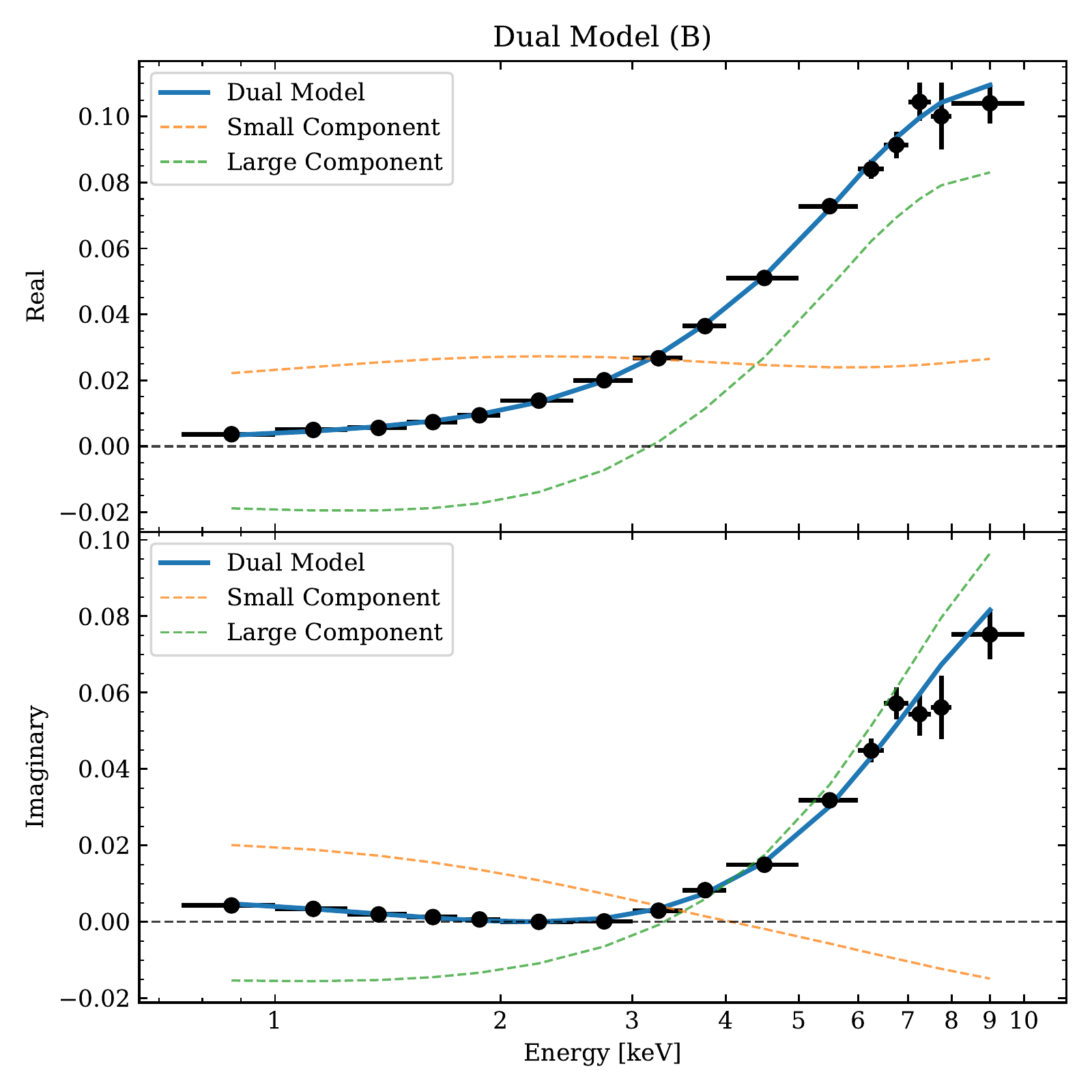}
  \caption{Same as Fig.~\ref{fig:fitting_dualA} for the best-fitting values to the energy-dependent fractional rms amplitude and phase lags of the Type-B QPO of \MAXI obtained with \dual model $B$.}
 \label{fig:fitting_dualB}
\end{figure*}

\begin{figure*}
  \includegraphics[angle=000,width=1.99\columnwidth]{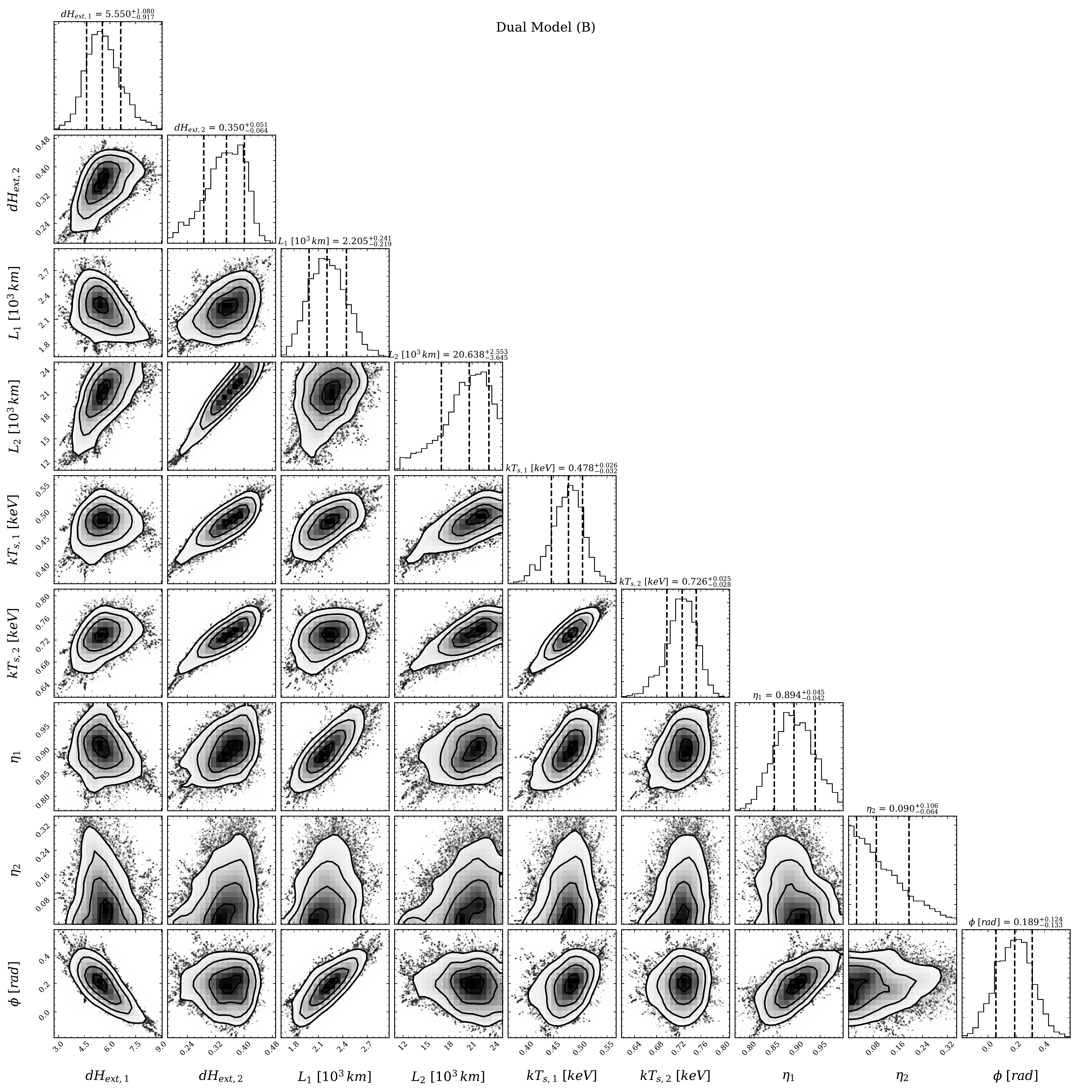}
  \caption{Same as Fig.~\ref{fig:corner_dualA} for the physical parameters of \MAXI obtained with \dual model $B$.}
 \label{fig:corner_dualB}
\end{figure*}

As described in Sec.~\ref{sec:dualmodel}, when considering the two-components {\em dual} model, we found two different solutions, namely Cases A and B. However, in Case B, the amplitudes of the temperature oscillations of one of the components are very high, and incompatible with the underlying perturbative assumptions of the model. Thus, for the sake of completeness, we left the full presentation of such Case to this Appendix. In Figure~\ref{fig:fitting_dualB} we show the energy-dependent fractional rms (upper-left panel) and phase lags (lower-left panel) of the type-B QPO in \MAXI, as fitted with this Case B best-fitting model. On the right panels we also plot the Real (upper panel) and Imaginary (bottom panel) parts of the complex oscillation, and the contribution of the individual components. In Figure~\ref{fig:corner_dualB} we show the corresponding corner plot for an MCMC chain of walkers concentrated around this solution. The best-fitting results and their uncertainties are presented in Table~\ref{tab:table_dualB}.

In this case, the {\em small} region has a size $L_1 \approx 2\,200$~km with high feedback, $\eta_1 = 0.89\pm0.04$, while the {\em large} region has $L_2 \approx 21\,000$~km and a small feedback, $\eta_2 < 0.22$ (consistent with zero). In this case, the soft-photon source temperatures are $kT_{s,1} \approx 0.5$ and $kT_{s,2} \approx 0.7$~keV. These temperatures are consistent with the temperature of the disc blackbody fitted to the soft-photon source in the time-averaged spectra, 
but rather high when the blackbody and disc blackbody spectra are compared.  

The physical properties of these two-components can be understood in the following framework: the small-sized corona ($L_1 \sim 150\,R_g$, assuming a 10~M$_\odot$ BH) fully surrounds the accretion disc ($kT_{s,1} \sim 0.5$~keV), impinging back into it a significant fraction of the inversely-Comptonised photons, and thus leading to a high feedback fraction ($\eta_1 \approx 0.9)$. The large Comptonisation region ($L_2 \sim 1500,R_g$), if spherical, should in principle be capable to produce strong feedback but, with such a large effective size, if elongated \cite[in a jet-like geometry, as possibly seen in \MAXI,][]{2019ATel12497....1C}, that would not be the case and the small feedback fraction ($\eta_2 < 0.2$) could be easily understood. 
Comptonisation in a precessing jet was invoked by \cite{2020A&A...640L..16K} to quantitatively describe the type-B QPO in GX~339--4. Moreover, as already suggested by \cite{2020MNRAS.496.4366B}, Comptonisation in a jet might be able to explain the spectral features of the type-B QPO in \MAXI. 
The relatively higher temperature of the soft-photon source associated to the large corona
in case B, could be explained if the photons injected were already partly-Comptonised in the small corona. 
However, the large amplitudes recovered for both temperatures, $\delta kT_s$ and $\delta kT_e$, and external heating rate, $\delta H_{\rm ext}$, of the {\em small} component, break the perturbative assumption in this Case.

\begin{table*}
 \caption{Best-fitting values for the physical parameters of \MAXI and their corresponding 1-$\sigma$ (68\%) uncertainties of the \dual model (Case B).}
 \label{tab:table_dualB}
 \begin{tabular}{ l | ccc | cccccc }
  \hline
  Component & $kT_e$ & $\Gamma$ & $\tau$ & $kT_s$ & $L$ & $\eta$ & ${\delta H}_{\rm ext}$ & $\delta kT_s$ & $\delta kT_e$\\
  & (keV) & & & (keV) & ($10^3$~km) & & (\%) & (\%) & (\%) \\
  \hline
  {\em Case B:} ~~ $\chi^2_\nu$ = 1.32 (23 dof) ~~ $\phi$ = 0.19$\pm$0.13 rad\\
  Small (1) & 20$^\dagger$ & 3.5$^\dagger$ & 1.3$^\dagger$ & 0.48$\pm$0.03 & 2.20$\pm$0.24 & 0.89$\pm$0.04 & 550$\pm$100 & 73$\pm$14 & 350$\pm$60 \\
  Large (2) & 20$^\dagger$ & 3.5$^\dagger$ & 1.3$^\dagger$ & 0.73$\pm$0.02 & 20.6$\pm$3.0 & <0.22 & 35$\pm$5 & <0.2 & 30$\pm$5\\
  \hline
  $^\dagger$ fixed parameters.
 \end{tabular}
\end{table*}

\bsp	
\label{lastpage}
\end{document}